\documentclass[12pt,a4paper]{iopart}

\expandafter\let\csname equation*\endcsname=\relax
\expandafter\let\csname endequation*\endcsname=\relax

\usepackage{amsmath,amsfonts,amsthm,amssymb}
\usepackage{color}
\usepackage{graphicx}
\usepackage{hyperref}
\usepackage{subcaption}
\usepackage{tikz}
\usepgflibrary{shapes.geometric}

\numberwithin{equation}{section}

\begin{document}

\title[Anomalous polymer collapse winding angle distributions]{Anomalous polymer collapse winding angle distributions\footnote{Dedicated to Professor Stu Whittington on the occasion of his 75th birthday}}
\author{A Narros$^1$, A L Owczarek$^2$ and T Prellberg$^1$}
\address{$^1$ School of Mathematical Sciences, Queen Mary University of London, Mile End Road, London E1 4NS, UK.}
\address{$^2$ School of Mathematics and Statistics,
  The University of Melbourne, Victoria 3010, Australia.}

\ead{owczarek@unimelb.edu.au, t.prellberg@qmul.ac.uk}

\date{\today}

\begin{abstract}
In two dimensions polymer collapse has been shown to be complex with multiple low temperature states and multi-critical points. Recently, strong numerical evidence has been provided for a long-standing prediction of universal scaling of winding angle distributions, where simulations of interacting self-avoiding walks show that the winding angle distribution for $N$-step walks is compatible with the theoretical prediction of a Gaussian with a variance growing asymptotically as $C\log N$. Here we extend this work by considering interacting self-avoiding trails which are believed to be a model representative of some of the more complex behaviour. We provide robust evidence that, while the high temperature swollen state of this model has a winding angle distribution that is also Gaussian, this breaks down at the polymer collapse point and at low temperatures. Moreover, we provide some evidence that the distributions are well modelled by stretched/compressed exponentials, in contradistinction to the behaviour found in interacting self-avoiding walks.

\end{abstract}

\section{Introduction}

Modelling the collapse of polymers in a dilute solution has advanced significantly in recent years with a variety of models demonstrating a range of  different behaviours that mimic some of the complexity seen in experiments. The classical theory of polymer collapse \cite{gennes1979a-a} has a high temperature swollen polymer becoming more compact as temperature is decreased until  a phase transition is reached at the so-called $\theta$-point, below which the polymer forms a liquid-like, or molten, globule. The $\theta$-point has been studied in both three and two dimensions where it is a critical point. The standard lattice model (see \cite{vanderzande1998a-a} for a review) is that of interacting self-avoiding walks which displays exactly this behaviour. For two dimensions see the extensive list of references in \cite{caracciolo2011a-a} including the key work by Duplantier and Saleur \cite{duplantier1987a-a}. However, it is known that polymers can also form folded or crystalline states, and various models have seen this type of phase with differing collapsing behaviour. One particular two-dimensional model that displays a different collapse critical behaviour is  the interacting self-avoiding trail model on a square lattice \cite{shapir1984a-a,owczarek1995a-:a,owczarek2006a-:a}, where the phase transition is much stronger than at the $\theta$-point in two dimensions and the low temperature state much more dense.

One interesting geometric property of polymer models is the winding angle distribution. A variety of models, including pure continuum Brownian motion and random walk models, have been studied \cite{spitzer1958a-a,rudnick1987a-a,belisle1989a-a,saleur1994a-a}. The two-dimensional self-avoiding walk is expected to have a winding angle distribution that is Gaussian \cite{fisher1984a-a,duplantier1988a-a} with variance proportional to $\log N$ ($N$ being the number of steps in the walk). More precisely,
\begin{equation}
P(x=\theta/\sqrt{\log N})\sim\exp(-x^2/4),
\end{equation}
where $\theta$ is the cumulative angle subtended by one end of the walk relative to the first step at the other end of the walk. This form implies that the variance of the winding angle distribution for swollen two-dimensional polymers behaves as 
\begin{equation}
\sigma^2\sim 2\log(N)\;.
\end{equation}
Using the theory of Coulomb gas \cite{duplantier1988a-a} leads to additional predictions for the $\theta$-point and collapsed two-dimensional polymers. The winding angle distribution remains Gaussian across the whole temperature range, with
\begin{equation}
P(x=\theta/\sqrt{\log N})\sim\exp(-x^2/(2C)),
\end{equation}
and universal values
 \begin{equation}
C=\begin{cases}2&\mbox{swollen phase,}\\ 24/7&\mbox{critical state,}\\ 4&\mbox{collapsed phase,}\end{cases}
\end{equation}
so  that
\begin{equation}
\sigma^2\sim C\log(N)\;.
\end{equation}

Recently,  this prediction has been supported by Monte Carlo studies of interacting self-avoiding walks up to length $N=400$ \cite{narros2016a-:a}.  It should be pointed out that earlier simulations of interacting self-avoiding walks up to length $N=300$ suggest that the results at the $\theta$-point and in the collapsed phase are more consistent with a stretched/compressed exponential of the type 
\begin{equation}
\label{stretched}
\exp(-|\theta|^\zeta/C\log N)
\end{equation}
 with $\zeta\approx1.5$ \cite{chang2000a-a}. As a consequence, the scaling variable would change to $x=\theta/(\log N)^{1/\zeta}$ and the variance would scale as
 \begin{equation}
 \label{variance}
\sigma^2\sim \left(C\log(N)\right)^{2/\zeta}.
\end{equation}
So it is clearly important that careful interpretation of numerical results be made.

Here we consider interacting self-avoiding trails since, as described above, they display different collapse behaviour. It is well known that pure self-avoiding trails, and so interacting self-avoiding trails (ISAT), at high temperature displays the same swollen polymer behaviour as self-avoiding walks. It is therefore of no surprise that our simulations for length up to $N=1000$ show that such trails have a Gaussian winding angle distribution with variance growing as $\log(N)$. On the other hand our data for the collapse point of ISAT, which is known exactly as $\beta=\beta_c \equiv \log(3)$, and for low temperatures is incompatible with a Gaussian winding angle distribution. Since we know the location of the collapse point exactly we can do a careful analysis of the point. We surprisingly find that the data is compatible with a compressed exponential distribution with an exponent $\zeta\approx 1.45$. We also have data for this point \cite{prellberg1998z-a} for trails of length $N=1,000,0000$ which supports this conclusion.
Given our own refutation \cite{narros2016a-:a} of the compressed exponential prediction for ISAW \cite{chang2000a-a} we are cautious about making a compressed exponential prediction for ISAT. It is appropriate to note here that a different compressed exponential prediction for ISAT was made in \cite{chang2000b-a} from simulations up to length $N=300$, where a value of $\zeta\approx 1.69$ was found.

\section{Results}

We performed ISAT simulations up to length $N=1000$ using the same method as described in our earlier work on ISAW \cite{narros2016a-:a}, based on the flatPERM algorithm proposed in \cite{prellberg2004a-a}. We also extended the ISAW simulations reported in \cite{narros2016a-:a} to length $N=1000$.

\begin{figure}[!hbt]
\begin{center}
\includegraphics[width=0.6\columnwidth]{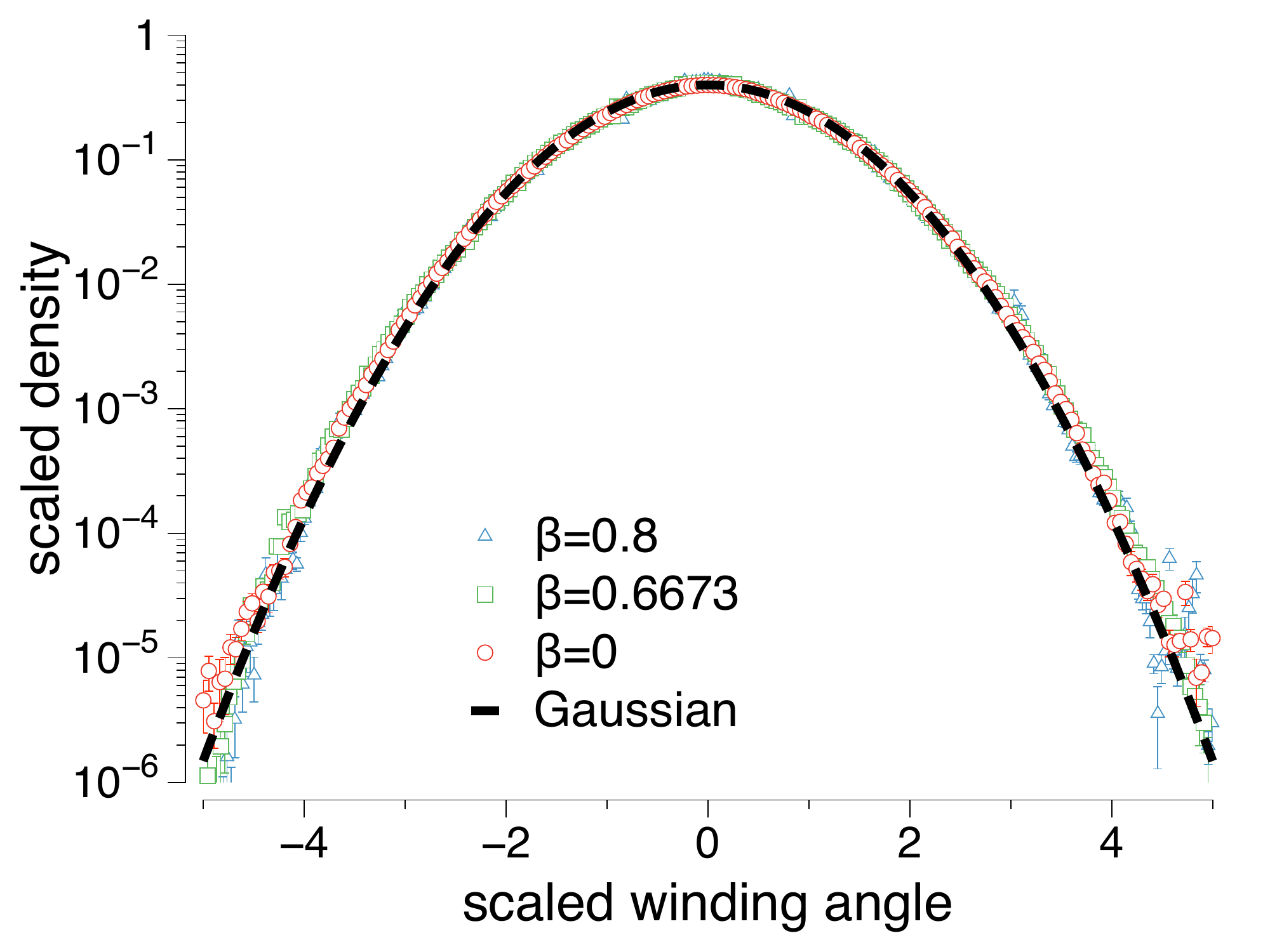}
\caption{The scaled winding angle distribution for the ISAW model of polymer collapse with $N=1000$ steps at $\beta=0$ (swollen phase), $\beta=0.6673$ ($\theta$-point), and $\beta=0.8$ (collapsed phase), demonstrating Gaussian behaviour at all temperatures.}
	\label{sec:res:fig1}
\end{center}
\end{figure}

To put our results for ISAT into context, in Figure \ref{sec:res:fig1} we show the agreement for winding angle distributions for ISAW with $N=1000$ steps in the swollen phase, $\theta$-point, and in the collapsed phase with a Gaussian distribution. Each of the distributions has been scaled to zero mean and unit variance, and there is no discernible deviation from a Gaussian distribution over nearly six orders of magnitude.

To probe deviations of a random variable from a normally distributed one, it is useful to consider its kurtosis. The kurtosis of a random variable $X$ with mean $\mu$ is the fourth standardised moment, defined as $\langle(X-\mu)^4\rangle/\langle(X-\mu)^2\rangle^2$, and the kurtosis of a normally distributed random variable is equal to $3$.

A finite-size extrapolation of the kurtosis of the winding angle distributions for ISAW at $\beta=0$, $\beta=0.6673$, and $\beta=0.8$  is shown in Figure \ref{sec:res:fig3}, and clearly shows that the numerical value of the kurtosis also approaches the Gaussian value of $3$ in the thermodynamic limit, assuming corrections proportional to $1/\log N$, which is the natural scaling for the variance as per Eqn.\ (\ref{variance}).

\begin{figure}[!hbt]
\begin{center}
\includegraphics[width=0.6\columnwidth]{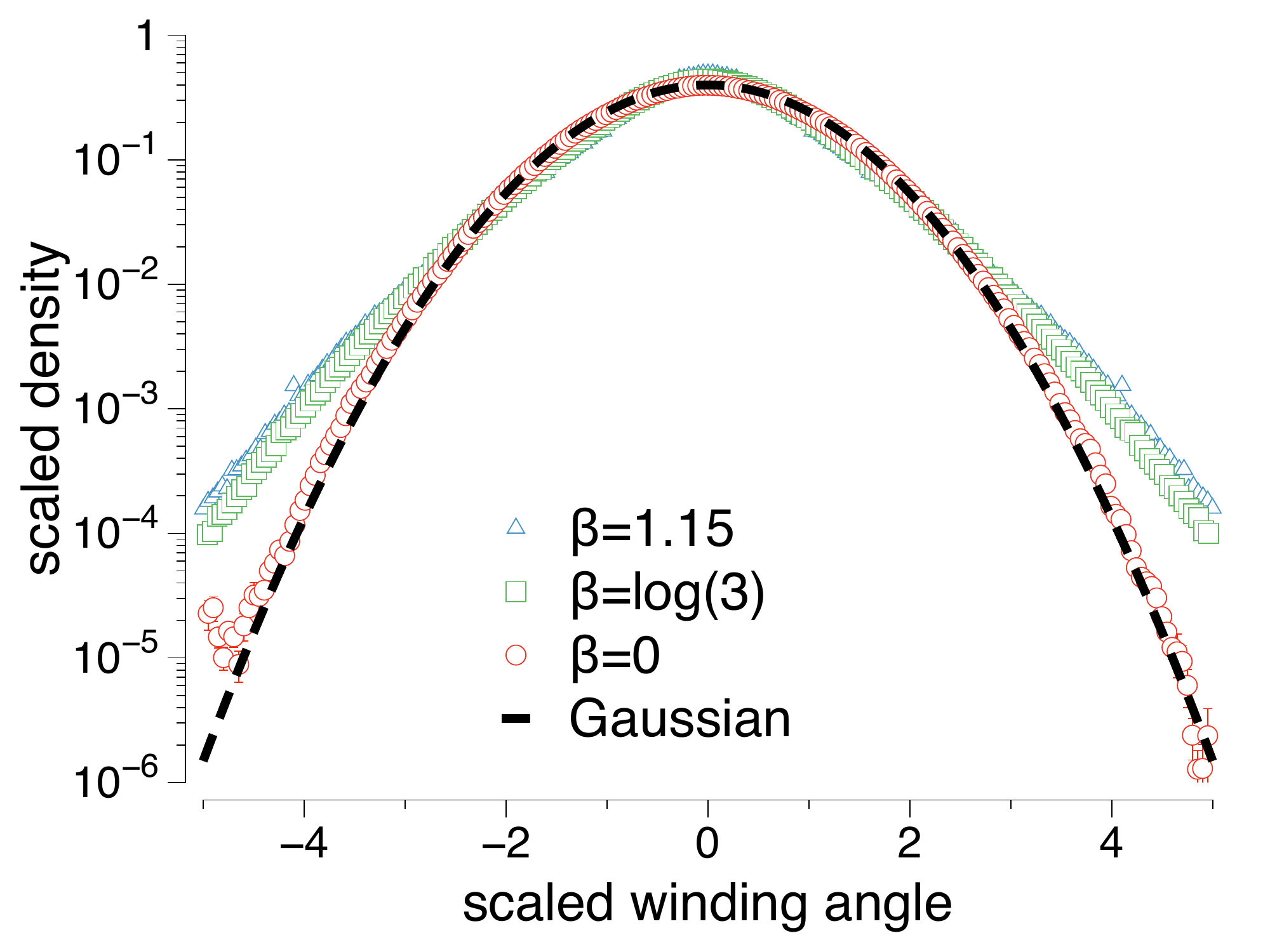}
\caption{The scaled winding angle distribution for the ISAT model of polymer collapse with $N=1000$ steps in the swollen phase, collapse point, and in the collapsed phase. While the swollen phase at $\beta=0$ demonstrates the Gaussian form, the collapse point at $\beta=\log(3)\approx 1.0986$ and the collapsed phase at low temperatures (here $\beta=1.15$) deviate substantially from this form.}
	\label{sec:res:fig2}
\end{center}
\end{figure}

We now move on to our new results for ISAT. Attempting to replicate the scaled distributions of Figure \ref{sec:res:fig1}, we show in Figure \ref{sec:res:fig2} winding angle distributions for $\beta=0$, $\beta=\log3$, and $\beta=1.15$, corresponding to the swollen phase of trails, critical trails, and collapsed trails, respectively. Once again, the distributions have been scaled to zero mean and unit variance. Only the distribution at $\beta=0$ is consistent with a Gaussian, and there are clear deviations from Gaussian behaviour for the two other distributions, which appear to be distinctly leptokurtic. 

\begin{figure}[!hbt]
\begin{center}
\includegraphics[width=0.75\columnwidth]{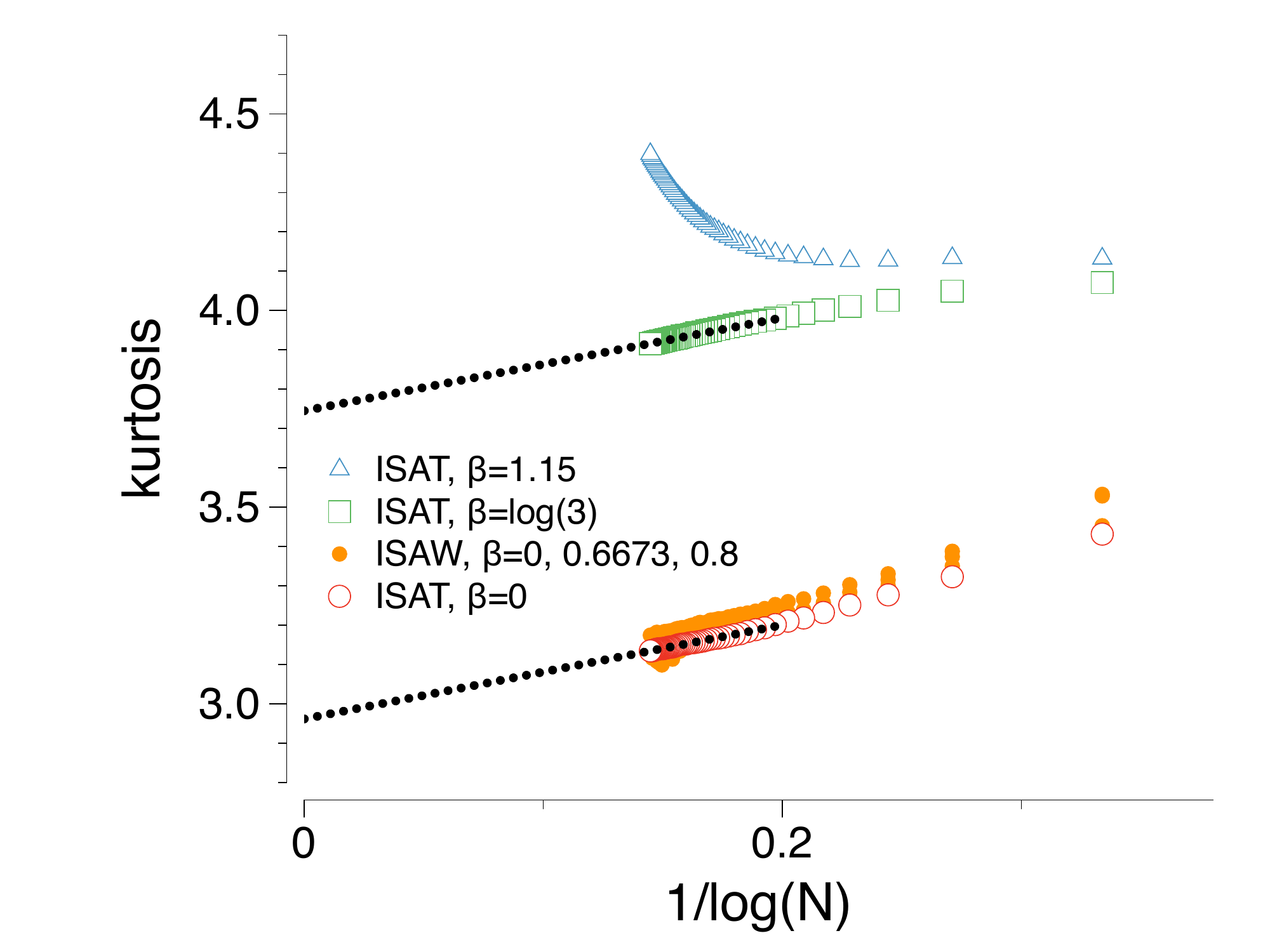}
\caption{The scaling of the kurtosis of the winding angle distribution of both ISAW and ISAT models at swollen ($\beta=0$), collapse point ($\beta=0.6673$ and $\beta=\log(3)$, respectively), and low temperatures ($\beta=0.8$ and $1.15$, respectively) plotted against $1/\log(N)$, for walks and trails from length $N=20$ to $N=1000$.  As expected all the ISAW data and the data for the swollen point of ISAT lie on top of each other and can be extrapolated to a value near the Gaussian value of $3$. However, it is very clear that at the collapse point of ISAT the kurtosis can be extrapolated on the same scale to around $3.76$ and is undoubtedly larger than $3$. At low temperatures, the kurtosis for ISAT \emph{increases} in length and is already larger than $4$ for the lengths considered in this work.}
	\label{sec:res:fig3}
\end{center}
\end{figure}

This naturally leads us to consider the finite-size extrapolation of the kurtosis of the winding angle distributions. Figure \ref{sec:res:fig3} shows that the kurtosis for ISAT at $\beta=0$ tends the Gaussian value of $3$ as was the case for ISAW at any temperature. However, while at the collapse point extrapolation against $1/\log N$ seems reasonable, the asymptotic value of the kurtosis is around $3.76$, clearly different from the Gaussian value. Even more striking is the low-temperature behaviour, where the kurtosis is greater than $4$ for all $N\geq20$ and is monotonically increasing with increasing $N$.

\begin{figure}[!hbt]
\begin{center}
\includegraphics[width=1\columnwidth]{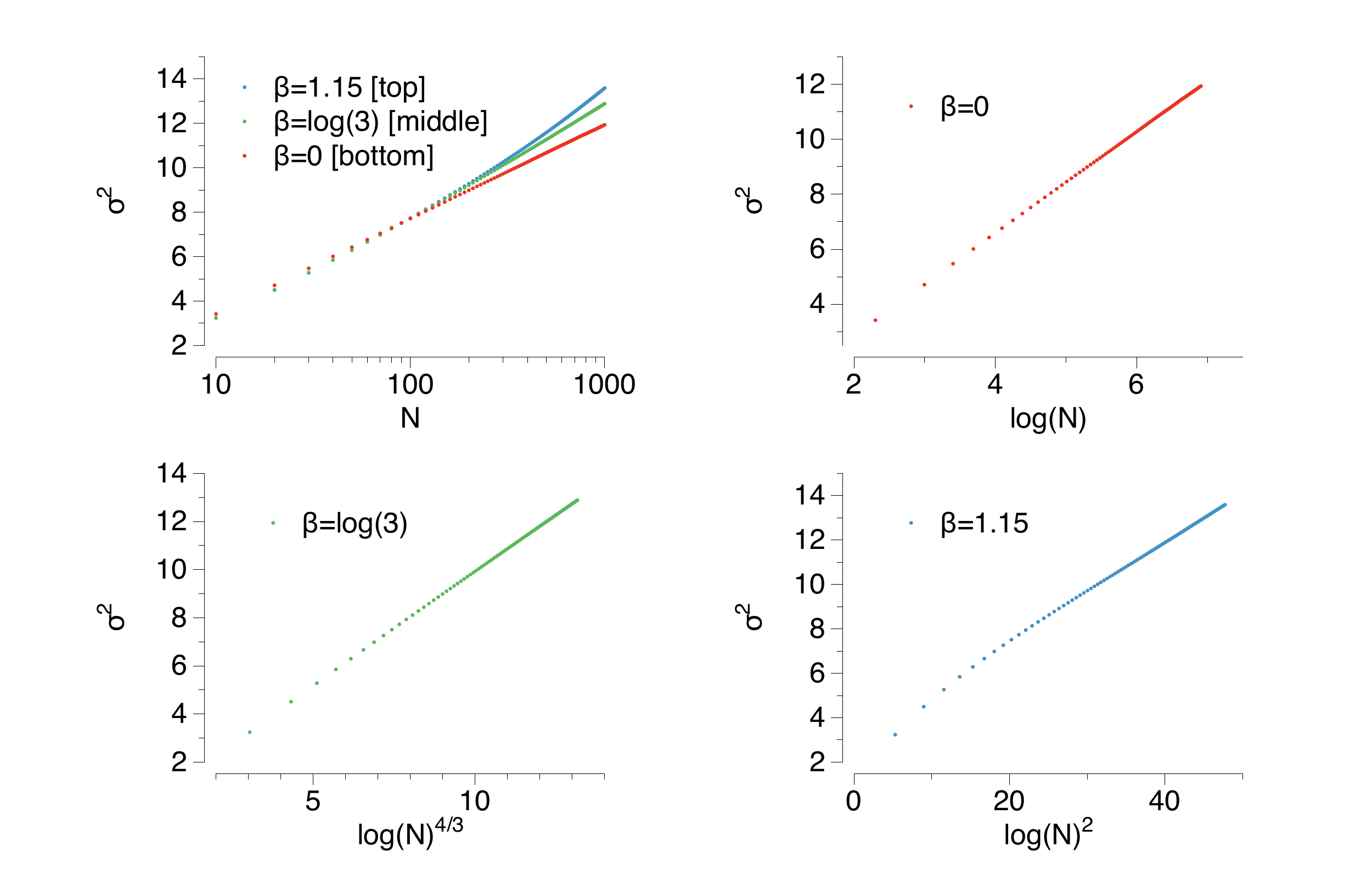}
\caption{
The four figures here show how the scaling of the variance of the winding angle distribution is linear in $\log(N)$ only for the swollen phase. In the top left the variance for all three temperatures at plotted on a semi-logarithm plot versus length $N$ for $N=10$ to $N=1000$. In the top right the variance is plotted against $\log(N)$ and clearly shows a linear behaviour reflecting the underlying Gaussian distribution. In the bottom left we plot the variance against $\log(N)^{4/3}$, which is a value in keeping with the asymptotic kurtosis value of $3.76$ found above. Finally, at the bottom right we plot the variance against $\log(N)^2$, which would be the case if the distribution was a pure linear exponential: we do not put too much credence on this but it is numerically plausible.
}
	\label{sec:res:fig4}
\end{center}
\end{figure}

We now turn to studying the growth of the winding angle variance in $N$. The top-left panel in Figure \ref{sec:res:fig4} shows that while the winding angle variance grows linearly in $\log N$ at $\beta=0$, there clearly is a deviation from linear growth at $\beta=\log 3$ and $\beta=1.15$. However, when the variance is plotted against suitably chosen powers of $\log N$, the growth appears again to be consistent with being roughly linear: against $\log N$ for $\beta=0$ (top-right), against $(\log N)^{4/3}$ for $\beta=\log 3$ (bottom-left), and against $(\log N)^2$ for $\beta=1.15$ (bottom-right).

To summarise the results so far, we have shown that at infinite temperature ($\beta=0$) the scaled winding angle distribution for ISAT is a Gaussian with kurtosis 3 and variance growing linearly in $\log N$. On the other hand, the winding angle distribution of critical ISAT ($\beta=\log 3$) is clearly different from Gaussian with kurtosis near $3.76$. Moreover, the distribution decays noticeably slower for large winding angles, and the variance grows faster than linearly in $\log N$. A simple heuristic mechanism for such a scenario is given by changing the scaling variable to $x=\theta/(\log N)^{1/\zeta}$. This leads to the stretched/compressed exponential distribution in Eqn.\ (\ref{stretched}). Assuming a compressed exponential of this form implies a variance growing as $(\log N)^{2/\zeta}$ and a value of the kurtosis given by
\begin{equation}
\frac{\Gamma(5/\zeta)\Gamma(1/\zeta)}{\Gamma(3/\zeta)^2}\;.
\end{equation}
For example, if $\zeta=3/2$ then the kurtosis is equal to $\frac{56}{81}\pi\sqrt3\approx3.76$. This would explain the rough linearity in the bottom-left panel of Figure \ref{sec:res:fig4}. Also note that a pure exponential, that is obtained by choosing $\zeta=1$, has a kurtosis of $6$. The behaviour of the variance at low temperatures as plotted in the bottom right panel of Figure \ref{sec:res:fig4}, which effectively assumes a pure exponential, is compatible with the behaviour of the kurtosis increasing dramatically in Figure \ref{sec:res:fig3}.

\begin{figure}[!hbt]
\begin{center}
\includegraphics[width=0.8\columnwidth]{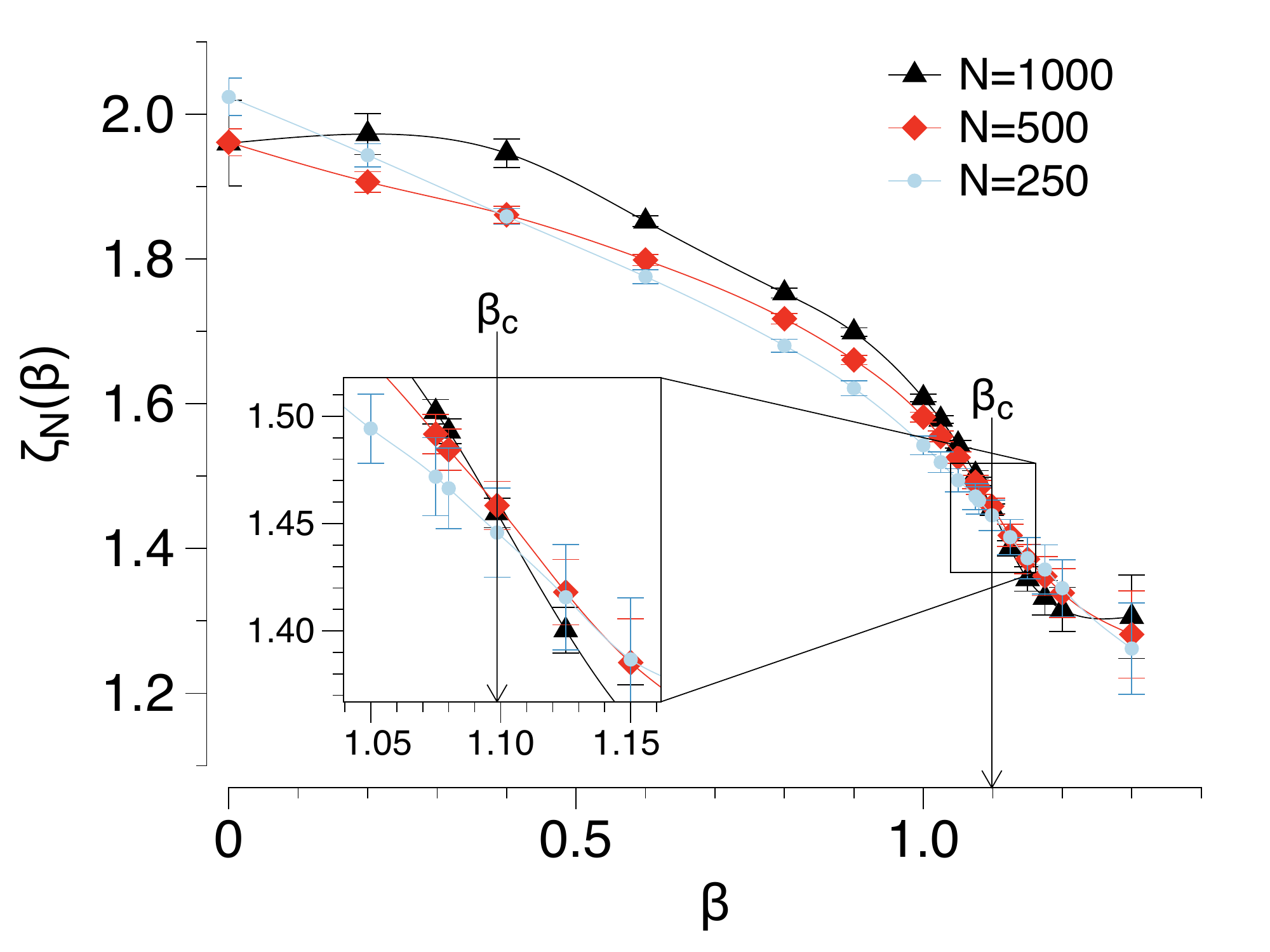}
\caption{A plot of the compressed exponential scaling exponent $\zeta_N(\beta) $  for three lengths $N=250, 500, 1000$ against inverse temperature $\beta$. The insert focuses on the critical temperature region. The three curves crosses near the critical temperature $\beta_c=\log(3)$ around an exponent value of $\zeta_c\approx 1.45$.
}
	\label{sec:res:fig5}
\end{center}
\end{figure}

The crossings of finite-size estimates of critical exponents are often used to determine the location of critical points. We apply this method to finite size estimates of the compressed exponential exponent, to see if the crossings correlate with the known critical temperature $\beta=\log 3$. If these crossings correlate then we can use the value of the compressed exponential exponent at the crossing point to confirm its existence and estimate its critical value. To do this, we fitted the winding angle distribution to the compressed exponential form given by Eqn.\ (\ref{stretched}) over a range of $\beta$ between $0$ and $1.4$ for lengths $N=250$, $500$, and $1000$. These estimates are plotted in Figure \ref{sec:res:fig5}. The estimates decrease from a value compatible with Gaussian behaviour at high temperature to lower values as temperature decreases, and indeed cross near the critical temperature $\beta=\log3\approx1.0986$. Assuming the critical temperature, we thus obtain an estimate of the critical value of $\zeta_c\approx1.45$. While we do not give error bars for $\zeta_c$, it would seem from the inset in Figure \ref{sec:res:fig5} that a value of $3/2$ is outside the probable range.

\begin{figure}[!hbt]
\begin{center}
\includegraphics[width=0.6\columnwidth]{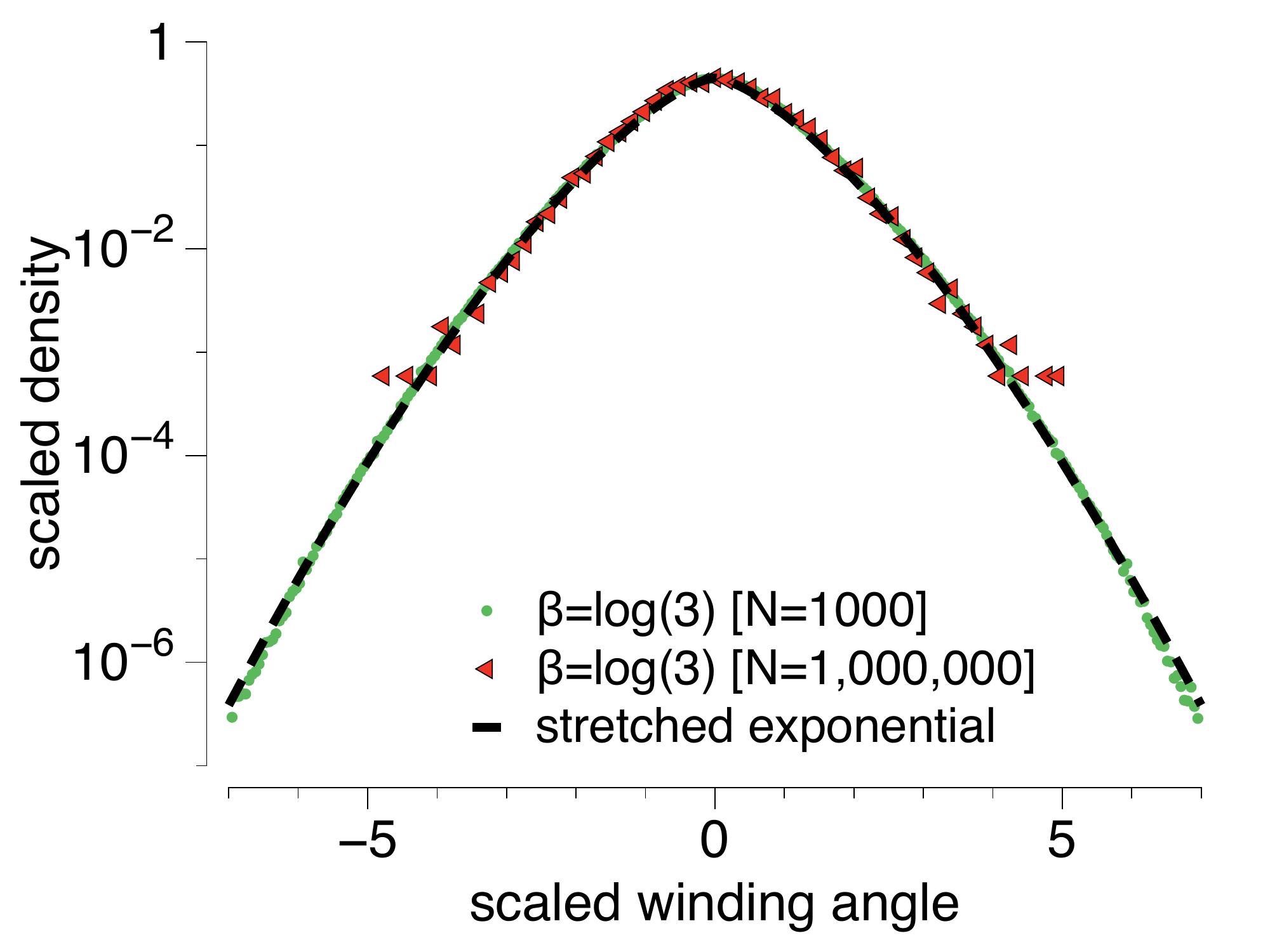}
\caption{The scaled winding angle distribution for ISAT with $N=1000$ at the collapse point $\beta=\log(3)$, with a fit based upon a compressed exponential with an exponent $\zeta =1.45$. The triangles (red online) show previously unpublished data \cite{prellberg1998z-a} for very long trails at length $N=10^6$ that are clearly compatible with this fit.
}
	\label{sec:res:fig6}
\end{center}
\end{figure}

We now show in Figure \ref{sec:res:fig6} the scaled winding angle distribution for ISAT with length $N=1000$ at $\beta=\log3$ and the compressed exponential distribution with the value $\zeta_c=1.45$, and find coincidence over six orders of magnitude. Additionally, we display simulation results for length $N=1,000,000$, albeit over a smaller range of scaled winding angles. Increasing the length of the walks by a factor of $1000$ does not seem to markedly shift the winding angle distribution from the compressed exponential form.

\begin{figure}[!hbt]
\begin{center}
\includegraphics[width=0.45\columnwidth]{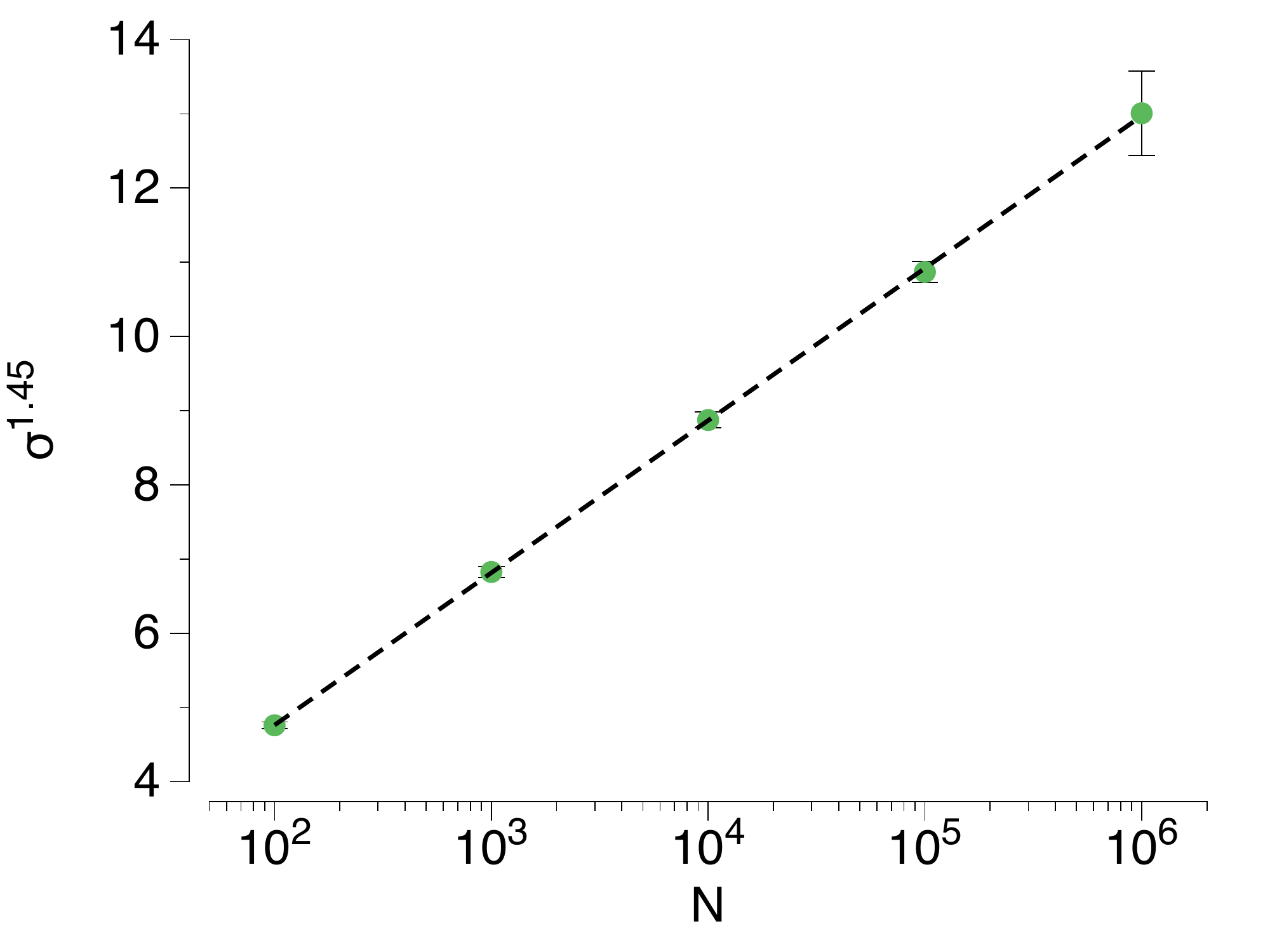}
\caption{Using previously unpublished data \cite{prellberg1998z-a} for very long trails at the collapse point up to length $N=10^6$, a scaled variance is plotted against length on a semi-logarithm scale for trails from length $N=100$ to $N=10^6$. This shows that the compressed exponential exponent of $1.45$ is strongly compatible with this longer data.
}
	\label{sec:res:fig7}
\end{center}
\end{figure}

Plotting the winding angle variance, raised to the estimated power $\zeta_c\approx1.45$,  against $\log N$ up to length $N=1,000,000$ shows strong compatibility across the full range, as shown in Figure \ref{sec:res:fig7}.

\section{Conclusions}
 
 We have studied the winding angle distribution of the interacting self-avoiding trail model of polymer collapse on the square lattice. This model has a collapse transition unlike the standard $\theta$-point and may represent a higher order multi-critical point in an enlarged parameter space. The nature of the collapsed phase also appears to be different to the standard molten globule. We provide strong evidence that while the high temperature swollen state of this model has a Gaussian winding angle distribution, the critical point and the low temperature phase do not. Moreover, we provide evidence that the collapse point is well modelled by a compressed exponential with an anomalous exponent $1.45$ rather than $2$ (the Gaussian value). Interestingly, this exponent value is close to, but not identical with the one observed in  a three-dimensional model of a polymer winding around a one-dimensional bar \cite{walter2011a-a}, for which the anomalous exponent was estimated to be $1.33(4)$ and the kurtosis was estimated as $3.74(5)$.


\ack

Financial support from the Australian Research
Council via its Discovery Projects scheme (DP160103562)
is gratefully acknowledged by one of the authors, A L Owczarek, who also 
thanks the School of Mathematical Sciences, Queen Mary University of London 
for hospitality. T Prellberg acknowledges support by EPSRC grant EP/L026708/1.

\section*{References}

\providecommand{\newblock}{}

\end{document}